\documentclass[10pt,twoside,a4paper]{article}

\pdfoutput=1

\usepackage{PICR2_with_page_numbers}

\newcommand{\beq}{\begin{equation}}
\newcommand{\eeq}{\end{equation}}
\newcommand{\beqa}{\begin{eqnarray}}
\newcommand{\eeqa}{\end{eqnarray}}
\newcommand{\matdd}[9]{\left(\begin{array}{ccc}{#1}&{#2}&{#3}\\%
                                               {#4}&{#5}&{#6}\\%
                                               {#7}&{#8}&{#9}%
                                               \end{array}\right)}
\newcommand{\bigfrac}[2]{\mbox{$\displaystyle\frac{#1}{#2}$}}

\newcommand{\pa}{\partial}
\newcommand{\D}{\mathrm{d}}
\newcommand{\grad}{\mbox{grad}\,}
\newcommand{\divg}{\mbox{div}\,}
\newcommand{\tr}{\mbox{tr}\,}
\newcommand{\sym}{\mbox{sym}\,}
\newcommand{\skw}{\mbox{skw}\,}
\newcommand{\pabl}[2]{\frac{\partial #1}{\partial #2}}
\newcommand{\bigpabl}[2]{\bigfrac{\partial #1}{\partial #2}}
\newcommand{\nl}{\nonumber\\}
\newcommand{\hf}{\mbox{$\frac{1}{2}$}}
\def\vecbu#1{\relax\ifmmode\mathchoice
  {\mbox{\boldmath$\bf\displaystyle#1$}}
  {\mbox{\boldmath$\bf\textstyle#1$}}
  {\mbox{\boldmath$\bf\scriptstyle#1$}}
  {\mbox{\boldmath$\bf\scriptscriptstyle#1$}}\else
  \hbox{\boldmath$\bf\textstyle#1$}\fi}
\def\tenssu#1{\relax\ifmmode\mathchoice
    {\mbox{$\sf\displaystyle#1$}}
    {\mbox{$\sf\textstyle#1$}}
    {\mbox{$\sf\scriptstyle#1$}}
    {\mbox{$\sf\scriptscriptstyle#1$}}\else
    \hbox{$\sf\textstyle#1$}\fi}

\begin{document}

\makeatletter
\newenvironment{figurehere}
  {\def\@captype{figure}}
  {}
\makeatother

\begin{center}
\pptitle{A Continuum-Mechanical Model for the Flow of
Anisotropic Polar Ice}

\ppauthor{Ralf Greve}{\ast}, \ppauthor{Luca Placidi}{\ast\ast},
\ppauthor{Hakime Seddik}{\ast}

\medskip
\ppaffil{\ast}{Institute of Low Temperature Science, Hokkaido
University, Sapporo, Japan; greve@lowtem.hokudai.ac.jp}
\ppaffil{\ast\ast}{Department of Structural and Geotechnical
Engineering, ``Sapienza'', University of Rome, Rome, Italy;
luca.placidi@uniroma1.it}

\bigskip
\end{center}

\vspace*{-44ex}

\noindent\parbox{\textwidth}{{\scriptsize\noindent{}\emph{Low
Temperature Science} \textbf{68}~(Suppl.), 137--148 (2009).
\\[-0.7ex]
\copyright{} Institute of Low Temperature Science, Hokkaido
University, Sapporo, Japan.}}

\vspace*{39ex}

\begin{multicols}{2}

\ppabstract{In order to study the mechanical behaviour of polar
ice masses, the method of continuum mechanics is used. The
newly developed CAFFE model (Continuum-mechanical, Anisotropic
Flow model, based on an anisotropic Flow Enhancement factor) is
described, which comprises an anisotropic flow law as well as a
fabric evolution equation. The flow law is an extension of the
isotropic Glen's flow law, in which anisotropy enters via an
enhancement factor that depends on the deformability of the
polycrystal. The fabric evolution equation results from an
orientational mass balance and includes constitutive relations
for grain rotation and recrystallization. The CAFFE model
fulfills all the fundamental principles of classical continuum
mechanics, is sufficiently simple to allow numerical
implementations in ice-flow models and contains only a limited
number of free parameters. The applicability of the CAFFE model
is demonstrated by a case study for the site of the EPICA
(European Project for Ice Coring in Antarctica) ice core in
Dronning Maud Land, East Antarctica.}

\ppkeyword{Ice, polycrystal, flow, anisotropy, continuum
mechanics, ice core}

\section{Introduction}

Ice in natural land ice masses, such as polar ice sheets, ice
caps or glaciers, consists of zillions of individual hexagonal
crystals (``crystallites'' or ``grains'') with a typical
diameter of millimeters to centimeters. This length scale
stands in contrast with the size of the ice masses, which
ranges from 100's of meters to 1000's of kilometers. It has
long been known that, while the distribution of the
crystallographic axes (also known as optical axes or $c$-axes)
at the surface of an ice sheet is essentially at random, deeper
down into the ice, different types of anisotropic fabrics with
preferred orientations of the $c$-axes tend to develop.

Many models have been proposed to describe the anisotropy of
polar ice. On the one end of the range in complexity, a simple
flow enhancement factor is introduced in an \emph{ad-hoc}
fashion as a multiplier of the isotropic ice fluidity in order
to account for anisotropy and/or impurities. This is done in
most current ice-sheet models, often without explicitly
mentioning anisotropy \cite{greve_05a, huybrechts_etal_07,
saito_abeouchi_04}. In macroscopic, phenomenological models, an
anisotropic macroscopic formulation for the flow law of the
polycrystal is postulated. To be usable, the rheological
parameters that enter this law must be evaluated as functions
of the anisotropic fabric \cite{gillet_chaulet_etal_05,
gillet_chaulet_etal_06, morland_staro_98, morland_staro_03}.
The concept of homogenization models, also called micro-macro
models, is to derive the polycrystalline behaviour at the level
of individual crystals and the fabric \cite{azuma_94,
castelnau_etal_98, castelnau_etal_96, goedert_hutter_98,
ktitarev_etal_02, lliboutry_93, svendsen_hutter_96,
thorsteinsson_02}. As for the ``high-end'' complexity,
full-field models solve the Stokes equation for ice flow
properly by decomposing the polycrystal into many elements,
which makes it possible to infer the stress and strain-rate
heterogeneities at the microscopic scale
\cite{lebensohn_etal_04a, lebensohn_etal_04b, mansuy_etal_02,
meyssonnier_philip_00}. A very comprehensive, up-to-date
overview of these different types of models is given by
Gagliardini et~al.\ \cite{gagliardini_etal_09} (in this
volume). However, the more sophisticated models are usually too
complex and computationally time-consuming to be included
readily in a model of macroscopic ice flow.

Here, the Continuum-mechanical, Anisotropic Flow model, based
on an anisotropic Flow Enhancement factor (``CAFFE model''),
shall be described. It belongs to the class of macroscopic
models, and is laid down in detail in the study by Placidi
et~al.\ \cite{placidi_etal_09} (based on previous works by
Faria \cite{faria_06a, faria_06b}, Faria et~al.\
\cite{faria_etal_06}, Placidi \cite{placidi_04, placidi_05},
Placidi and Hutter \cite{placidi_hutter_06a}). The flow
enhancement factor is taken as a function of a newly introduced
scalar quantity called \emph{deformability}, which is
essentially a non-dimensional invariant related to the shear
stress acting on the basal plane of a crystallite, weighted by
the orientation-distribution function which describes the
anisotropic fabric of the polycrystal. Fabric evolution is
modelled by an orientation mass balance which accounts for
grain rotation and recrystallization processes.

The CAFFE model fulfills all the fundamental principles of
classical continuum mechanics (see also Placidi et~al.\
\cite{placidi_hutter_06b}), and it is a good compromise between
necessary simplicity on the one hand, and consideration of the
major effects of anisotropy on the other. In order to
demonstrate its performance, the model is applied to the site
of the EPICA (European Project for Ice Coring in Antarctica)
ice core in Dronning Maud Land, East Antarctica, for which data
on the ice flow as well as on the anisotropic fabric are
available.

\section{CAFFE model}

\subsection{Glen's flow law}
\label{sect_glen}

\begin{table*}[htb]
  \centering
  \begin{tabular}{ll} \hline
  \textbf{Quantity}\rule{0mm}{2.3ex} & \textbf{Value}\\ \hline
  Stress exponent, $n$\rule{0mm}{2.25ex} & $3$ \\
  Pre-exponential constant, $A_0$\rule{0mm}{2.5ex} &
  \hspace*{-0.8em} $\begin{array}[t]{ll}
    3.985\times{}10^{-13}\,\mathrm{s^{-1}\,Pa^{-3}}
    & \mbox{(for $T'\le{}263.15\,\mathrm{K}$)}
    \\
    1.916\times{}10^{3}\,\mathrm{s^{-1}\,Pa^{-3}}
    & \mbox{(for $T'>263.15\,\mathrm{K}$)}
  \end{array}$ \\
  Activation energy, $Q$\rule{0mm}{2.5ex} &
  \hspace*{-0.8em} $\begin{array}[t]{ll}
    60\,\mathrm{kJ\,mol^{-1}}
    & \hspace*{4.4em} \mbox{(for $T'\le{}263.15\,\mathrm{K}$)}
    \\
    139\,\mathrm{kJ\,mol^{-1}}
    & \hspace*{4.4em} \mbox{(for $T'>263.15\,\mathrm{K}$)}
  \end{array}$ \\ \hline
  \end{tabular}
  \caption{\emph{Physical parameters for Glen's flow law.}}
  \label{tab_physical_parameters}
\end{table*}

Let us briefly review the isotropic case, for which
polycrystalline ice is treated as an incompressible, viscous
fluid. The Cauchy stress tensor $\tenssu{T}$ is split up
according to
\beq
  \tenssu{T} = -p\tenssu{1} + \tenssu{S},
  \qquad
  p = -\frac{1}{3} \tr\tenssu{T},
  \label{eq_stress_deviator}
\eeq
where $p$ denotes the pressure, and $\tenssu{S}$ is the
traceless stress deviator ($\tr\tenssu{S}=0$). Due to the
incompressibility, the flow law only determines the stress
deviator $\tenssu{S}$ and reads
\beq
  \tenssu{S} = 2\eta\tenssu{D},
  \label{eq_viscous_flow_law}
\eeq
where $\tenssu{D}=\sym\grad\vecbu{v}$ is the strain-rate tensor
(symmetric part of the gradient of the velocity $\vecbu{v}$), and
the coefficient $\eta$ is the \emph{shear viscosity} (or simply the
\emph{viscosity}). Its inverse, the \emph{fluidity}, can be
factorized as
\beq
  \frac{1}{\eta} = 2EA(T')f(\sigma),
  \label{eq_fluidity}
\eeq
where
\beq
  \sigma = \sqrt{\hf\,\tr(\tenssu{S}^2)}
  \label{eq_effective_stress}
\eeq
is the \emph{effective stress} (square root of the second invariant
of the stress deviator), and the \emph{creep function} $f(\sigma)$
is given by the power law
\beq
  f(\sigma) = \sigma^{n-1}
  \label{eq_power_law}
\eeq
(the parameter $n$ is called ``stress exponent''). Further, the
\emph{rate factor} $A(T')$ depends on the temperature relative
to pressure melting $T'=T-T_\mathrm{m}+T_0$ ($T$: absolute
temperature, $T_\mathrm{m}=T_0-\beta{}p\,$: pressure melting
point, $T_0=273.16\,\mathrm{K}\,$: melting point at zero
pressure, $\beta=9.8\times{}10^{-2}\,\mathrm{K\,MPa^{-1}}$:
Clausius-Clapeyron constant) via the Arrhenius law
\beq
  A(T') = A_0\,e^{-Q/RT'},
  \label{eq_arrhenius_law}
\eeq
where $A_0$ is the pre-exponential constant, $Q$ the activation
energy and $R=8.314\,\mathrm{J\,mol^{-1}\,K^{-1}}$ the
universal gas constant. The \emph{flow enhancement factor} $E$
is equal to unity for pure ice, and can be set to values
deviating from unity in order to account roughly for effects of
impurities and/or anisotropy (Paterson \cite{paterson_91}).

The isotropic flow law for ice is now obtained by inserting
Eq.~(\ref{eq_fluidity}) [with the specifications of
Eqs.~(\ref{eq_power_law}) and (\ref{eq_arrhenius_law})] in the
viscous flow law (\ref{eq_viscous_flow_law}). This yields
\beq
  \tenssu{D} = EA(T')f(\sigma) \, \tenssu{S},
  \label{eq_glens_flow_law}
\eeq
which is called \emph{Nye's generalization of Glen's flow law},
or \emph{Glen's flow law} for short (e.g., Greve and Blatter
\cite{greve_blatter_09}, Hooke \cite{hooke_05}, Paterson
\cite{paterson_94}, van der Veen \cite{vanderveen_99}).
Suitable values for the several parameters are listed in
Table~\ref{tab_physical_parameters}.

\subsection{Anisotropic generalization of Glen's flow law}

\subsubsection{Deformation of a crystallite}

In order to derive a generalization of Glen's flow law
(\ref{eq_glens_flow_law}) which accounts for general,
anisotropic fabrics of the ice polycrystal, we first consider
the deformation of a crystallite embedded in the
polycrystalline aggregate. Following Placidi et~al.\
\cite{placidi_etal_09}, only the dominant deformation along the
basal plane is accounted for, whereas deformations along
prismatic and pyramidal planes, which are at least 60 times
more difficult to activate, shall be neglected
(Fig.~\ref{fig_glide_planes.pdf}).

\begin{figurehere}
  \medskip
  \centering
  \noindent\includegraphics[width=75mm]{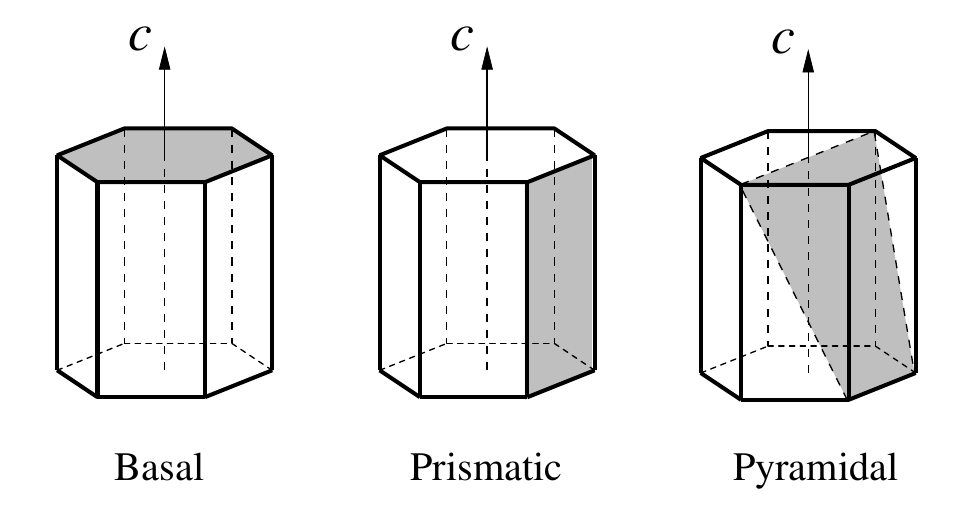}
  \par\vspace*{-2mm}\par
  \caption{\emph{Basal, prismatic and pyramidal glide planes in the
  hexagonal ice crystal, sketched as a right hexagonal prism
  (Faria \cite{faria_03}).}}
  \label{fig_glide_planes.pdf}
  \medskip
\end{figurehere}

Let $\vecbu{n}$ be the normal unit vector of the basal plane
(direction of the $c$-axis), then $\tenssu{T}\vecbu{n}$ is the
resolved stress vector (Fig.~\ref{fig_stress_decomposition}).
Note that the tensor $\tenssu{T}$ is interpreted as the
\emph{macroscopic} stress which does not depend on the
orientation $\vecbu{n}$. It is reasonable to assume that only
the stress component $S_\mathrm{t}$ tangential to the basal
plane (resolved shear stress) contributes to its shear
deformation, while the component normal to the basal plane has
no effect.

\begin{figurehere}
  \medskip
  \centering
  \noindent\includegraphics[width=45mm]{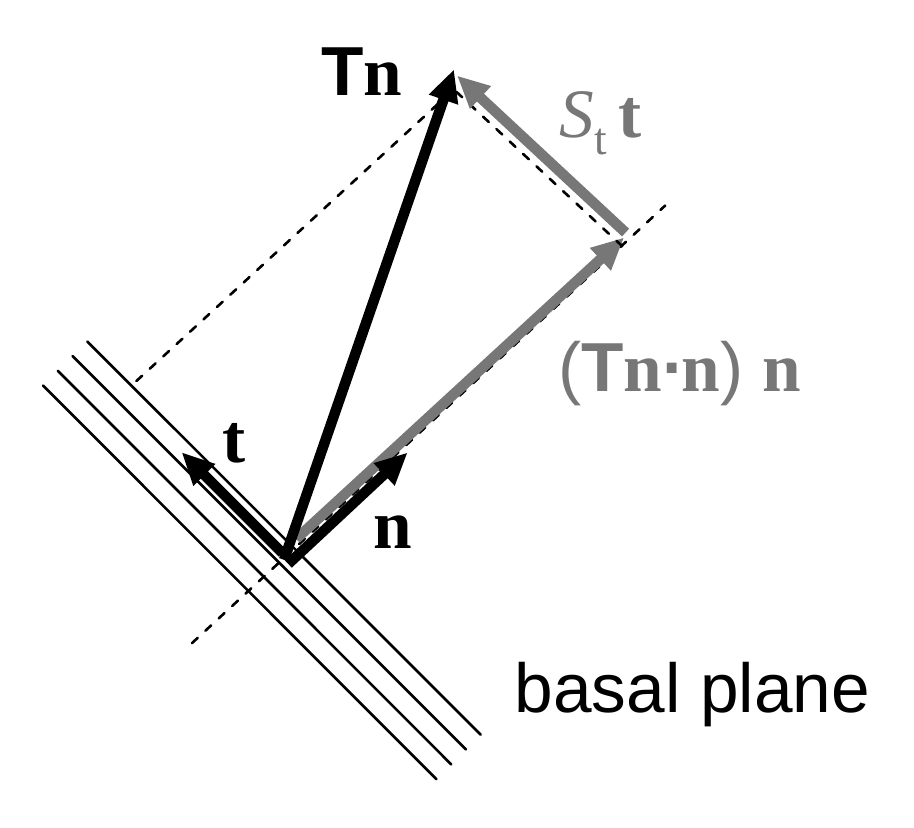}
  \par\vspace*{-2mm}\par
  \caption{\emph{Decomposition of the stress vector into a part
  normal and a part tangential to the basal plane.}}
  \label{fig_stress_decomposition}
  \medskip
\end{figurehere}

According to Fig.~\ref{fig_stress_decomposition}, the decomposition
of the stress vector reads
\beq
  \tenssu{T}\vecbu{n}
  = (\tenssu{T}\vecbu{n}\cdot\vecbu{n})\vecbu{n}
    + S_\mathrm{t}\vecbu{t},
  \label{eq_stress_decomposition_1}
\eeq
where $\vecbu{t}$ denotes the tangential unit vector. Inserting
the decomposition (\ref{eq_stress_deviator}) of the stress
tensor $\tenssu{T}$ readily eliminates the pressure $p$ and
leaves
\beq
  \tenssu{S}\vecbu{n}
  = (\tenssu{S}\vecbu{n}\cdot\vecbu{n})\vecbu{n}
    + S_\mathrm{t}\vecbu{t}.
  \label{eq_stress_decomposition}
\eeq
As mentioned above, deformation of the crystallite in the
polycrystalline aggregate is attributed to the tangential
component $S_\mathrm{t}$ only. Since we aim at a theory which
describes the effects of anisotropy by a \emph{scalar},
anisotropic flow enhancement factor, we define the scalar
invariant
\beq
  S_\mathrm{t}^2
  = \tenssu{S}\vecbu{n}\cdot\tenssu{S}\vecbu{n}
    - (\tenssu{S}\vecbu{n}\cdot\vecbu{n})^2.
  \label{eq_sc_st}
\eeq
This quantity has the unit of a stress squared, and so a
natural way to non-dimensionalize it is by the square of the
effective stress $\sigma$ [Eq.~(\ref{eq_effective_stress})],
which is also a scalar invariant. Thus, we introduce the
\emph{deformability} of a crystallite in the polycrystalline
aggregate, which is loaded by the stress $\tenssu{T}$, as
\beq
  \mathcal{A}^\star(\vecbu{n})
    = \frac{5}{2}\,\frac{S_\mathrm{t}^2(\vecbu{n})}{\sigma^2}
    = 5\,\frac{\tenssu{S}\vecbu{n}\cdot\tenssu{S}\vecbu{n}
               -(\tenssu{S}\vecbu{n}\cdot\vecbu{n})^2}
              {\tr(\tenssu{S}^2)}.
  \label{eq_deformability_crystallite}
\eeq
The factor $5/2$ has been introduced merely for reasons of
convenience, as it will become clear below.

\subsubsection{Flow law for polycrystalline ice}

In polycrystalline ice, the crystallites within a control
volume (which is assumed to be large compared to the
crystallite dimensions, but small compared to the macroscopic
scale of ice flow) show a certain fabric. Extreme cases are on
the one hand the single maximum fabric, for which all $c$-axes
are perfectly aligned, and on the other hand the isotropic
fabric with a completely random distribution of the $c$-axes. A
general fabric, which is usually in between these cases, can be
described by the \emph{orientation mass density} (OMD)
$\rho^\star(\vecbu{n})$. It is defined as the mass per volume
and orientation, the latter being specified by the normal unit
vector (direction of the $c$-axis) $\vecbu{n}\in{}S^2$ ($S^2$
is the unit sphere). Evidently, when integrated over all
orientations, the OMD must yield the normal mass density
$\rho$, which leads to the normalization condition
\beq
  \int\limits_{S^2} \rho^\star(\vecbu{n})\,\D^2 n = \rho.
  \label{eq_omd_normalization}
\eeq
Alternatively, an \emph{orientation distribution function} (ODF)
$f^\star(\vecbu{n})$ can be defined as
\beq
  f^\star(\vecbu{n}) = \frac{\rho^\star(\vecbu{n})}{\rho},
  \label{eq_odf}
\eeq
which is normalized to unity when integrated over all orientations.

We use the ODF in order to define the \emph{deformability} of
polycrystalline ice by weighting the deformability of the
crystallite (\ref{eq_deformability_crystallite}),
\beqa
  \mathcal{A}
  &\!\!\!=\!\!\!& \int\limits_{S^2}
      \mathcal{A}^\star(\vecbu{n})\,f^\star(\vecbu{n})\,\D^2 n
  \nl[1ex]
  &\!\!\!=\!\!\!& \frac{5}{2} \int\limits_{S^2}
      \frac{S_\mathrm{t}^2(\vecbu{n})}{\sigma^2}
      f^\star(\vecbu{n})\,\D^2 n
  \nl[1ex]
  &\!\!\!=\!\!\!& 5 \int\limits_{S^2}
      \frac{\tenssu{S}\vecbu{n}\cdot\tenssu{S}\vecbu{n}
            -(\tenssu{S}\vecbu{n}\cdot\vecbu{n})^2}
           {\tr(\tenssu{S}^2)}
      f^\star(\vecbu{n})\,\D^2 n.
  \label{eq_deformability}
\eeqa
Note that, for the isotropic case, the ODF is
$f^\star(\vecbu{n})=1/(4\pi)$, and from
Eq.~(\ref{eq_deformability}) we obtain a deformability of
$\mathcal{A}=1$ (Placidi et~al.\ \cite{placidi_etal_09}). For
that reason, the factor $5/2$ has been introduced in
Eq.~(\ref{eq_deformability_crystallite}).

\begin{figurehere}
  \medskip
  \centering
  \noindent\includegraphics[width=50mm]{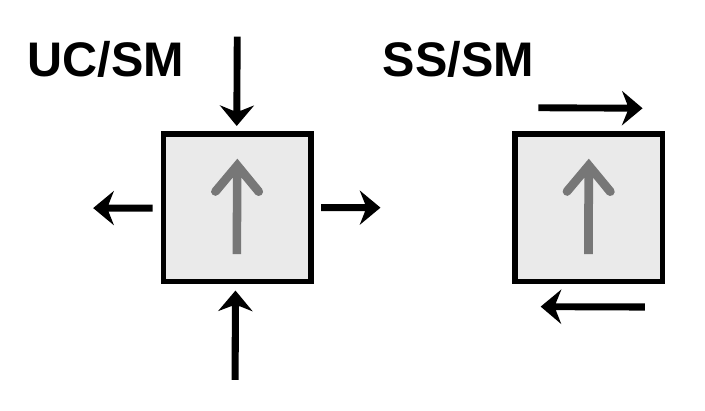}
  \caption{\emph{Uniaxial compression on single maximum \emph{(UC/SM)}
  and simple shear on single maximum \emph{(SS/SM)} for a small sample
  of polycrystalline ice. Stresses are indicated as black arrows,
  and the single maximum fabric is marked by the dark-grey arrows
  within the ice sample.}}
  \label{fig_stress_states}
  \medskip
\end{figurehere}

The envisaged flow law for anisotropic polar ice can now be
formulated. Essentially, we keep the form of the Glen's flow law
(\ref{eq_glens_flow_law}), but with a scalar, anisotropic
enhancement factor $\hat{E}(\mathcal{A})$ instead of the parameter
$E$,
\beq
  \tenssu{D} = \hat{E}(\mathcal{A})\,A(T')f(\sigma) \, \tenssu{S}.
  \label{eq_glens_flow_law_aniso}
\eeq
The function $\hat{E}(\mathcal{A})$ is supposed to be strictly
increasing with the deformability $\mathcal{A}$, and has the fixed
points
\beq
  \begin{array}{l}
    \hat{E}(0) = E_\mathrm{min}
    \\[1ex]
    \;\;\;\mbox{(uniaxial compression on single maximum)},
    \\[1ex]
    \hat{E}(1) = 1
    \\[1ex]
    \;\;\;\mbox{(arbitrary stress on isotropic fabric)},
    \\[1ex]
    \hat{E}(\frac{5}{2}) = E_\mathrm{max}
    \\[1ex]
    \;\;\;\mbox{(simple shear on single maximum)}.
  \end{array}
  \label{eq_aniso_enh_factor_values}
\eeq
The ``hard'' case (\ref{eq_aniso_enh_factor_values})$_1$ and
the ``soft'' case (\ref{eq_aniso_enh_factor_values})$_3$ are
illustrated in Fig.~\ref{fig_stress_states}. Note also that the
deformability cannot take values larger than $\mathcal{A}=5/2$
(Placidi et~al.\ \cite{placidi_etal_09}).

For the detailed form $\hat{E}(\mathcal{A})$ of the anisotropic
enhancement factor, in addition to
Eq.~(\ref{eq_aniso_enh_factor_values}), we demand that the
function is continuously differentiable, that is,
$\hat{E}\in{}C^1[0,\frac{5}{2}]$. Moreover, Azuma
\cite{azuma_95} and Miyamoto \cite{miyamoto_99} have
experimentally verified that the enhancement factor depends on
the Schmid factor (resolved shear stress) to the fourth power,
that is, on the square of the deformability $\mathcal{A}$. This
yields
\beq
  \hat{E}(\mathcal{A}) = \left\{
  \begin{array}{l}
    E_\mathrm{min} + (1-E_\mathrm{min}) \mathcal{A}^t,
    \\[3ex]
    \hspace*{3.5em}
    t = \bigfrac{8}{21}\,\bigfrac{E_\mathrm{max}-1}{1-E_\mathrm{min}},
    \\[3ex]
    \hspace*{9em}
    0 \le \mathcal{A} \le 1,
    \\[3ex]
    \bigfrac{4\mathcal{A}^2(E_\mathrm{max}-1)+25-4E_\mathrm{max}}{21},
    \\[3ex]
    \hspace*{9em}
    1 \le \mathcal{A} \le \bigfrac{5}{2}
  \end{array} \right.
  \label{eq_aniso_enh_factor}
\eeq
(for details see Placidi et~al.\ \cite{placidi_etal_09}).
Several investigations (e.g.\ Budd and Jacka
\cite{budd_jacka_89}, Pimienta et~al.\ \cite{pimienta_etal_87},
Russell-Head and Budd \cite{russell_head_budd_79}) indicate
that the parameter $E_\mathrm{max}$ (maximum softening) is
approximately equal to ten. The parameter $E_\mathrm{min}$
(maximum hardening) can be realistically chosen between zero
and one tenth, a non-zero value serving mainly the purpose of
avoiding numerical problems. The function
(\ref{eq_aniso_enh_factor}) is shown in
Fig.~\ref{fig_aniso_enh_factor}.

\begin{figurehere}
  \medskip
  \centering
  \noindent\includegraphics[width=60mm]{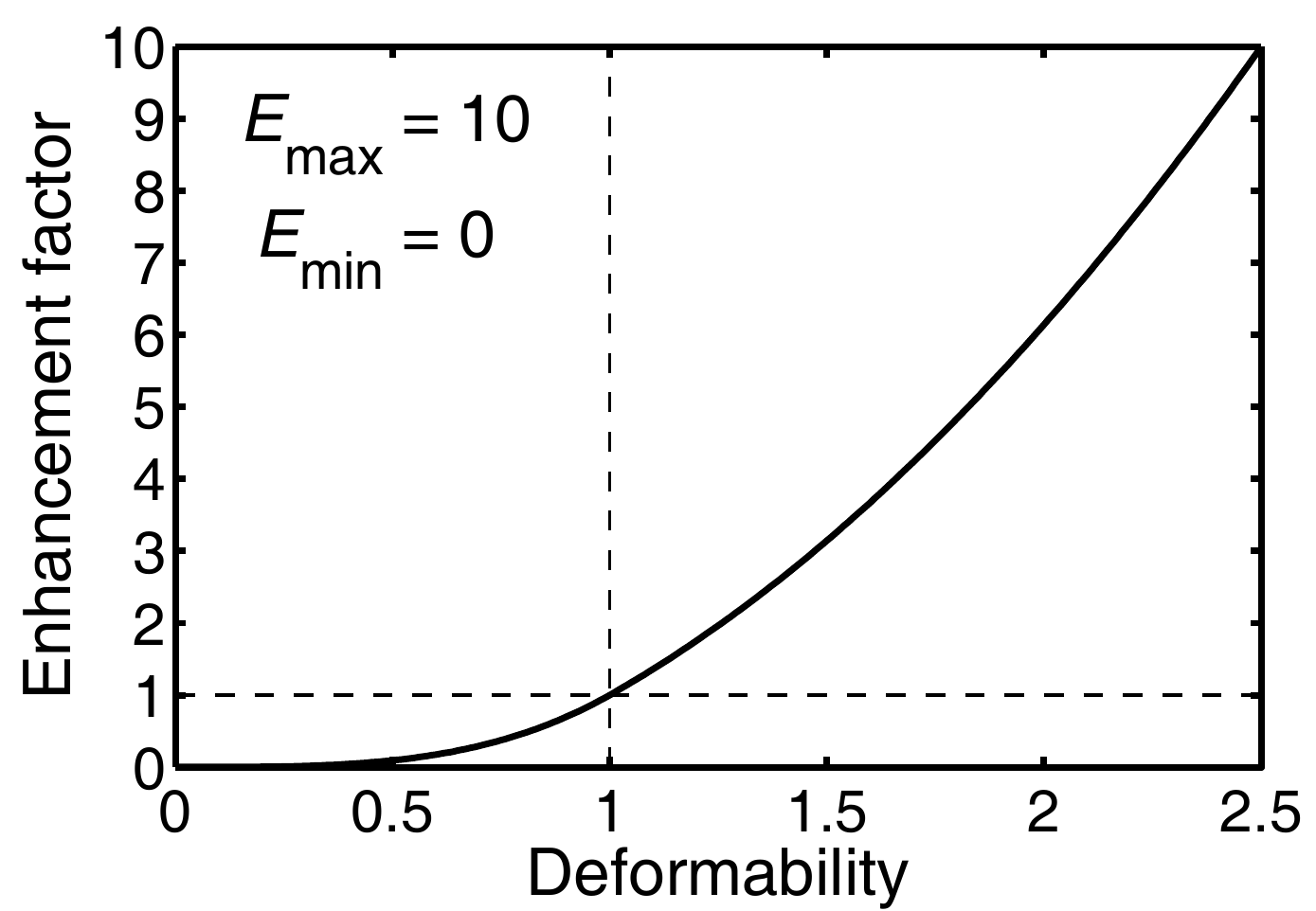}
  \caption{\emph{Anisotropic enhancement factor $\hat{E}(\mathcal{A})$
  as a function of the deformability $\mathcal{A}$ according to
  Eq.~(\ref{eq_aniso_enh_factor}), for $E_\mathrm{max}=10$ and
  $E_\mathrm{min}=0$.}}
  \label{fig_aniso_enh_factor}
  \medskip
\end{figurehere}

\subsubsection{Inversion of the flow law}

As long as the creep function $f(\sigma)$ is given by the power law
(\ref{eq_power_law}), the anisotropic flow law
(\ref{eq_glens_flow_law_aniso}) can be inverted analytically. We
find
\beq
  \tenssu{S} = [\hat{E}(\mathcal{A})]^{-1/n}\,[A(T')]^{-1/n} \,
               d^{-(1-1/n)} \, \tenssu{D},
  \label{eq_glens_flow_law_aniso_inv}
\eeq
where
\beq
  d = \sqrt{\hf\,\tr(\tenssu{D}^2)}
  \label{eq_effective_strain_rate}
\eeq
is the \emph{effective strain rate}. The deformability
$\mathcal{A}$ also needs to be expressed by strain rates
instead of stresses [see Eq.~(\ref{eq_deformability})]. In
analogy to Eq.~(\ref{eq_stress_decomposition}), we consider the
resolved strain-rate vector $\tenssu{D}\vecbu{n}$ in a
crystallite in the polycrystalline aggregate, and decompose it
according to
\beq
  \tenssu{D}\vecbu{n}
  = (\tenssu{D}\vecbu{n}\cdot\vecbu{n})\vecbu{n}
    + D_\mathrm{t}\vecbu{t},
  \label{eq_strain_rate_decomposition}
\eeq
where $D_\mathrm{t}$ is the resolved shear rate in the basal
plane (see also Fig.~\ref{fig_stress_decomposition}). As in
Eq.~(\ref{eq_sc_st}), we define the scalar invariant
\beq
  D_\mathrm{t}^2
  = \tenssu{D}\vecbu{n}\cdot\tenssu{D}\vecbu{n}
    - (\tenssu{D}\vecbu{n}\cdot\vecbu{n})^2.
  \label{eq_dc_dt}
\eeq
Owing to the collinearity of the tensors $\tenssu{S}$ and
$\tenssu{D}$ [see Eqs.~(\ref{eq_glens_flow_law_aniso}) and
(\ref{eq_glens_flow_law_aniso_inv})], the deformability of a
crystallite in the polycrystalline aggregate
[Eq.~(\ref{eq_deformability_crystallite})] can be readily
expressed by $D_\mathrm{t}$ and $d$,
\beq
  \mathcal{A}^\star(\vecbu{n})
    = \frac{5}{2}\,\frac{D_\mathrm{t}^2(\vecbu{n})}{d^2}
    = 5\,\frac{\tenssu{D}\vecbu{n}\cdot\tenssu{D}\vecbu{n}
               -(\tenssu{D}\vecbu{n}\cdot\vecbu{n})^2}
              {\tr(\tenssu{D}^2)},
  \label{eq_deformability_crystallite_inv}
\eeq
and the deformability of polycrystalline ice
[Eq.~(\ref{eq_deformability})] yields
\beqa
  \mathcal{A}
  &\!\!\!=\!\!\!& \int\limits_{S^2}
      \mathcal{A}^\star(\vecbu{n})\,f^\star(\vecbu{n})\,\D^2 n
  \nl[1ex]
  &\!\!\!=\!\!\!& \frac{5}{2} \int\limits_{S^2}
      \frac{D_\mathrm{t}^2(\vecbu{n})}{d^2}
      f^\star(\vecbu{n})\,\D^2 n
  \nl[1ex]
  &\!\!\!=\!\!\!& 5 \int\limits_{S^2}
      \frac{\tenssu{D}\vecbu{n}\cdot\tenssu{D}\vecbu{n}
            -(\tenssu{D}\vecbu{n}\cdot\vecbu{n})^2}
           {\tr(\tenssu{D}^2)}
      f^\star(\vecbu{n})\,\D^2 n.
  \label{eq_deformability_inv}
\eeqa
This completes the inversion of the anisotropic flow law.

\subsection{Evolution of anisotropy}

\subsubsection{Orientation mass balance}

The anisotropic flow law in the form
(\ref{eq_glens_flow_law_aniso}) or
(\ref{eq_glens_flow_law_aniso_inv}) needs to be complemented by
an evolution equation for the anisotropic fabric. This is done
by formulating the \emph{orientation mass balance} for the OMD
$\rho^\star(\vecbu{n})$.

\par\rule{0mm}{5mm}\par

\begin{figurehere}
  \medskip
  \centering
  \noindent\includegraphics[width=55mm]{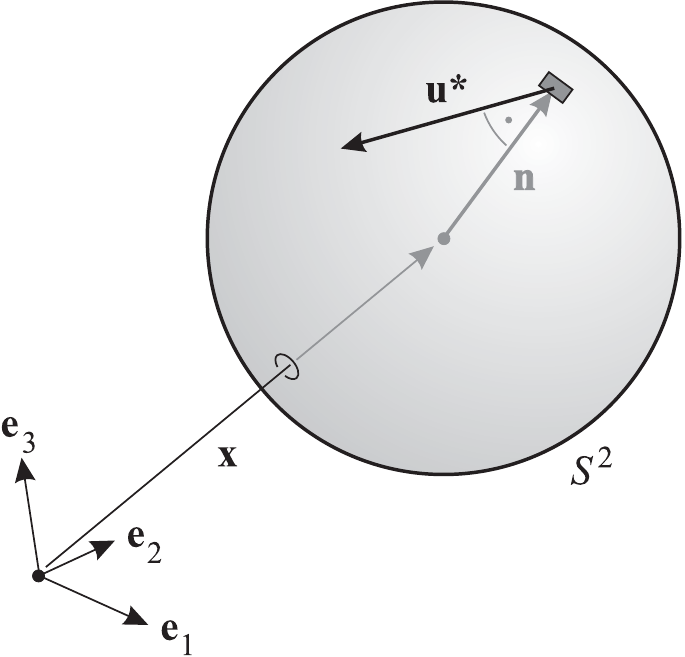}
  \caption{\emph{Orientation transition rate
  $\vecbu{u}^\star(\vecbu{n})$ on the unit sphere $S^2$.}}
  \label{fig_otr}
  \medskip
\end{figurehere}

We are not going to enter the detailed formalism of orientation
balance equations here (see Placidi et~al.\
\cite{placidi_etal_09}, and references therein). Instead, we
rather motivate the form of the orientation mass balance by
generalizing the ordinary mass balance. The difference is that,
in addition to the dependencies on the position vector
$\vecbu{x}$ and the time $t$, the density and velocity fields
also depend on the orientation vector $\vecbu{n}\in{}S^2$,
which is indicated by the notation $\rho^\star(\vecbu{n})$ and
$\vecbu{v}^\star(\vecbu{n})$. The velocity, which describes
motions in the physical space, is complemented by an
\emph{orientation transition rate}
$\vecbu{u}^\star(\vecbu{n})$, which describes motions on the
unit sphere, that is, changes of the orientation due to grain
rotation (Fig.~\ref{fig_otr}). Also, an \emph{orientation flux}
$\vecbu{q}^\star(\vecbu{n})$ is considered, which allows
redistributions of the OMD due to rotation recrystallization
(polygonization). Consequently, the orientation mass balance
reads
\beq
  \pabl{\rho^\star}{t} + \divg(\rho^\star\vecbu{v}^\star)
   + \divg\!_{S^2}(\rho^\star\vecbu{u}^\star+\vecbu{q}^\star)
  = \rho^\star \Gamma^\star.
  \label{eq_omb_1}
\eeq
The first two terms on the left-hand side are straightforward
generalizations of the terms in the ordinary mass balance. The
third term on the left-hand side is the equivalent of the
second term for the orientation transition rate
$\vecbu{u}^\star(\vecbu{n})$ and the orientation flux
$\vecbu{q}^\star(\vecbu{n})$, where $\divg\!_{S^2}$ is the
divergence operator on the unit sphere. On the right-hand side,
a source term appears which allows that certain orientations
can be produced at the expense of others. The quantity
$\Gamma^\star(\vecbu{n})$ is therefore called the
\emph{orientation production rate}. Physically, it describes
migration recrystallization and all other processes in which
the transport of mass from one grain, having a certain
orientation, to another grain, having a different orientation,
cannot be neglected.

\par\vspace*{2ex}\par

In the following, we will make the reasonable assumption that
the spatial velocity does not depend on the orientation, that
is, $\vecbu{v}^\star(\vecbu{n})=\vecbu{v}$. Therefore, the
orientation mass balance (\ref{eq_omb_1}) simplifies to
\beq
  \pabl{\rho^\star}{t} + \divg(\rho^\star\vecbu{v})
   + \divg\!_{S^2}(\rho^\star\vecbu{u}^\star+\vecbu{q}^\star)
  = \rho^\star \Gamma^\star.
  \label{eq_omb}
\eeq
Integration over $S^2$ (all orientations) gives the classical
mass balance
\beq
  \pabl{\rho}{t} + \divg(\rho\vecbu{v}) = 0,
  \label{eq_mass_balance}
\eeq
with the use of the Gauss theorem and the mass-conservation
requirement
\beq
  \int\limits_{S^2} \rho^\star(\vecbu{n})\,\Gamma^\star(\vecbu{n})
  \,\D^2 n = 0.
  \label{eq_mass_conservation}
\eeq

\par\vspace*{1ex}\par

In order to solve the orientation mass balance (\ref{eq_omb}),
constitutive relations for the orientation transition rate
$\vecbu{u}^\star(\vecbu{n})$, the orientation flux
$\vecbu{q}^\star(\vecbu{n})$ and the orientation production
rate $\Gamma^\star(\vecbu{n})$ need to be provided as closure
conditions.

\par\vspace*{2.5ex}\par

\subsubsection{Constitutive relation for the orientation transition
rate}

As mentioned above, the orientation transition rate corresponds
physically to grain rotation. Since grain rotation is induced
by shear deformation in the basal plane, we argue that it is
controlled by the resolved shear rate $D_\mathrm{t}\vecbu{t}$
[Eq.~(\ref{eq_strain_rate_decomposition})], and use the
relation
\beqa
  \vecbu{u}^\star(\vecbu{n})
  &=& -\iota \, D_\mathrm{t}\vecbu{t} + \tenssu{W}\vecbu{n}
  \nl[0.5ex]
  &=& \iota \cdot
      [(\tenssu{D}\vecbu{n}\cdot\vecbu{n})\vecbu{n}-\tenssu{D}\vecbu{n}]
      + \tenssu{W}\vecbu{n}
  \label{eq_otr}
\eeqa
(see, e.g., Dafalias \cite{dafalias_01}). The parameter $\iota$
is assumed to be a positive constant. The additional term
$\tenssu{W}\vecbu{n}$ with the spin tensor
$\tenssu{W}=\skw\grad\vecbu{v}$ (skew-symmetric part of the
gradient of the velocity $\vecbu{v}$) describes the
contribution of local rigid-body rotations.

In the special case $\iota=1$, the basal planes are material
area elements, that is, they carry out an affine rotation.
However, due to geometric incompatibilities of the deformation
of individual crystallites in the polycrystalline aggregate, an
affine rotation is not plausible, and we expect realistic
values of $\iota$ to be less than unity.

\subsubsection{Constitutive relation for the orientation flux}

The orientation flux is supposed to describe rotation
recrystallization (polygonization). Following the argumentation
by G\"odert \cite{goedert_03}, it is modelled as a diffusive
process,
\beq
  \vecbu{q}^\star(\vecbu{n})
  = -\lambda\,\grad\!_{S^2}
     [\rho^\star(\vecbu{n})\mathcal{H}^\star(\vecbu{n})],
  \label{eq_ofl}
\eeq
where the parameter $\lambda>0$ is the orientation diffusivity,
$\grad\!_{S^2}$ is the gradient operator on the unit sphere,
and the ``hardness'' $\mathcal{H}^\star(\vecbu{n})$ is a
monotonically decreasing function of the crystallite
deformability $\mathcal{A}^\star(\vecbu{n})$ [see
Eq.~(\ref{eq_deformability_crystallite})]. A simple choice for
the hardness function would therefore be
\beq
  \mathcal{H}^\star(\vecbu{n})
  = \frac{1}{\mathcal{A}^\star(\vecbu{n})+\epsilon},
  \label{eq_hardness_1}
\eeq
the offset $\epsilon\ll{}1$ being introduced in order to
prevent a singularity for $\mathcal{A}^\star=0$. However,
recent results by Durand et~al.\ \cite{durand_etal_08} suggest
that rotation recrystallization is an isotropic process not
affected by the orientation. In this case, the choice
\beq
  \mathcal{H}^\star(\vecbu{n}) \equiv 1
  \label{eq_hardness_2}
\eeq
is indicated, which renders Eq.~(\ref{eq_ofl}) equivalent to
Fick's laws of diffusion on the unit sphere.

\subsubsection{Constitutive relation for the orientation production
rate}

The driving force for the orientation production rate, which
models essentially migration recrystallization, are macroscopic
deformations of the polycrystal, which can be more easily
followed on the microscopic scale by grains oriented favourably
for the given deformation. Therefore, it is reasonable to
assume that the orientation production rate for a certain
orientation $\vecbu{n}$ is related to the crystallite
deformability $\mathcal{A}^\star(\vecbu{n})$
[Eqs.~(\ref{eq_deformability_crystallite}),
(\ref{eq_deformability_crystallite_inv})]. In the CAFFE model,
the linear relation
\beq
  \Gamma^\star(\vecbu{n})
  = \Gamma\,[\mathcal{A}^\star(\vecbu{n})-\mathcal{A}]
  \label{eq_opr}
\eeq
is proposed. Subtraction of the polycrystal deformability
$\mathcal{A}$ is required in order to fulfill the
mass-conservation condition (\ref{eq_mass_conservation}). The
parameter $\Gamma$ is assumed to be positive, which guarantees
a positive mass production for favourably oriented grains, and
a negative production for unfavourably oriented grains
(Fig.~\ref{fig_opr}).

\begin{figurehere}
  \medskip
  \centering
  \noindent\includegraphics[width=55mm]{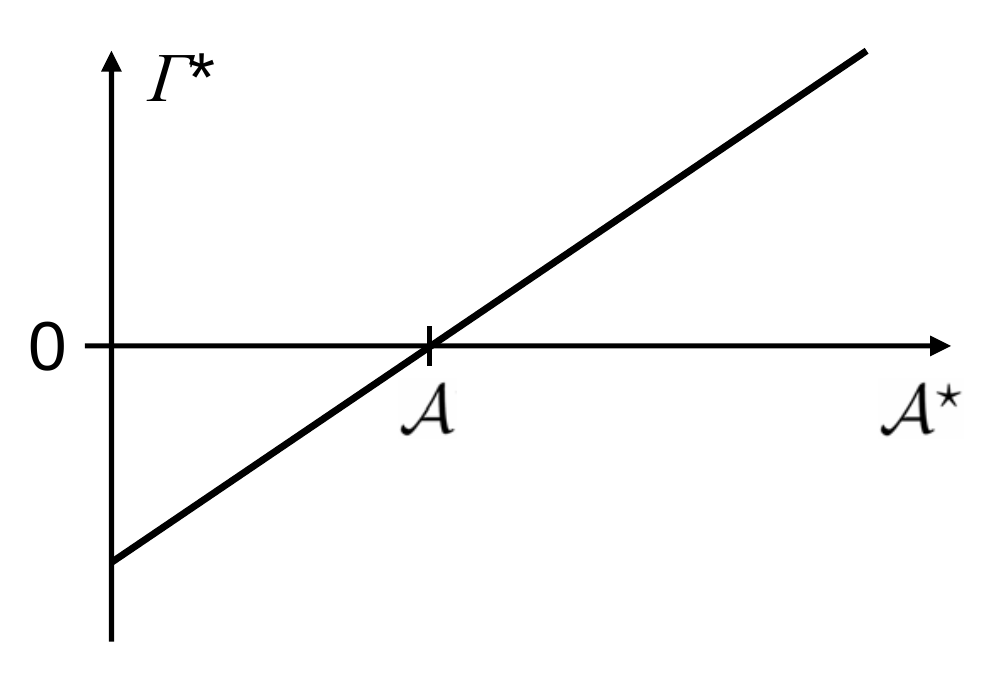}
  \caption{\emph{Orientation production rate according to
  Eq.~(\ref{eq_opr}).}}
  \label{fig_opr}
  \medskip
\end{figurehere}

The CAFFE model is now formulated completely. Equation
(\ref{eq_glens_flow_law_aniso}) is the actual flow law, which
replaces its isotropic counterpart (\ref{eq_glens_flow_law}).
Anisotropy enters via the enhancement factor
$\hat{E}(\mathcal{A})$ [Eq.~(\ref{eq_aniso_enh_factor})], which
depends on the deformability $\mathcal{A}$ defined in
Eq.~(\ref{eq_deformability}). Computation of the deformability
requires knowledge of the orientation mass density
$\rho^\star$, which is governed by the evolution equation
(\ref{eq_omb}) and the constitutive relations (\ref{eq_otr}),
(\ref{eq_ofl}) and (\ref{eq_opr}).

\section{Application to the EDML ice core}

\subsection{Methods}

We have developed a one-dimensional flow model, including the
CAFFE model, for the site of the EPICA ice core at Kohnen
Station in Dronning Maud Land, East Antarctica (``EDML core'',
$75^\circ{}00'06''\mathrm{S}$, $00^\circ{}04'04''\mathrm{E}$,
2892 meters above sea level; see
http://www.awi-bremerhaven.de/Polar/Kohnen/). For this core
with an overall length of 2774~m, preliminary fabric data are
available from 50~m until 2570~m depth (I.\ Hamann, pers.\
comm.\ 2006). The fabric is essentially isotropic down to
approximately 600~m depth, and shows a gradual transition to a
broad girdle fabric between 600 and 1000~m depth. Further down,
the girdle fabric narrows until approximately 2000~m depth. The
fabric then experiences an abrupt change towards a single
maximum, which prevails below 2040~m depth. Tendencies of
secondary or multiple maxima are observed at several depths.
The complete data set and a detailed interpretation will be
presented elsewhere (Hamann et~al.\ \cite{hamann_etal_09}).

The location of the EDML site on a flank (rather than a dome
like most other ice cores) allows deriving a one-dimensional
flow model based on the shallow-ice approximation (Hutter
\cite{hutter_83}, Morland \cite{morland_84}), with which the
performance of the CAFFE model can be tested. We define a local
Cartesian coordinate system such that Kohnen Station is located
at the origin, the $x$-axis points in the $260^\circ$ (WSW)
direction, the $y$-axis in the $170^\circ$ (SSE) direction, and
$z$ (depth) points vertically downward
(Fig.~\ref{fig_coordinate_system}).

\begin{figurehere}
  \medskip
  \centering
  \noindent\includegraphics[width=60mm]{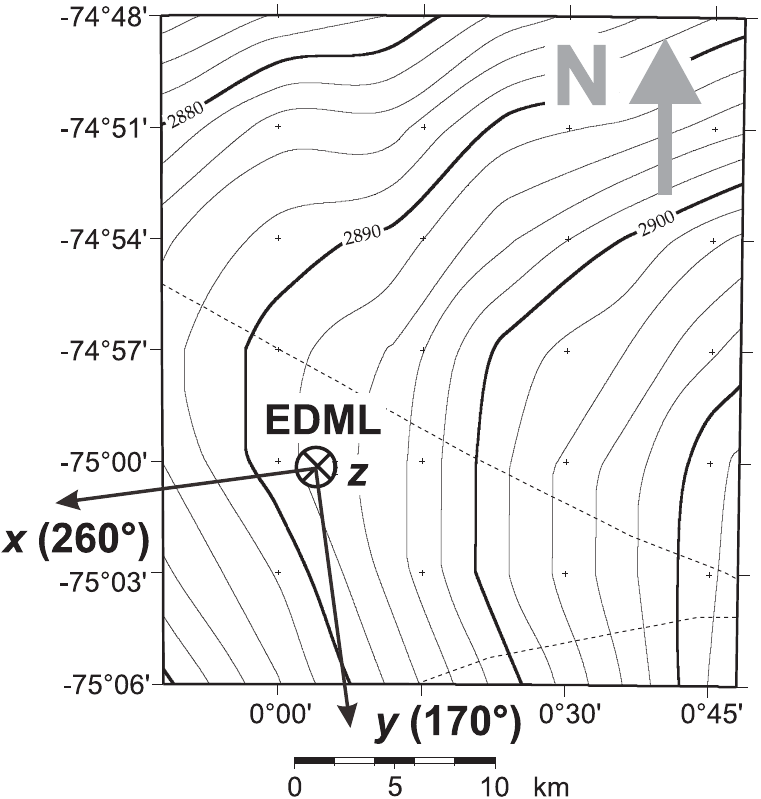}
  \caption{\emph{Local coordinate system for the EDML site. Underlaid
  topography map by Wesche et~al.\ \cite{wesche_etal_07}.}}
  \label{fig_coordinate_system}
  \medskip
\end{figurehere}

\begin{figure*}[htb]
  \centering
  \includegraphics[width=110mm]{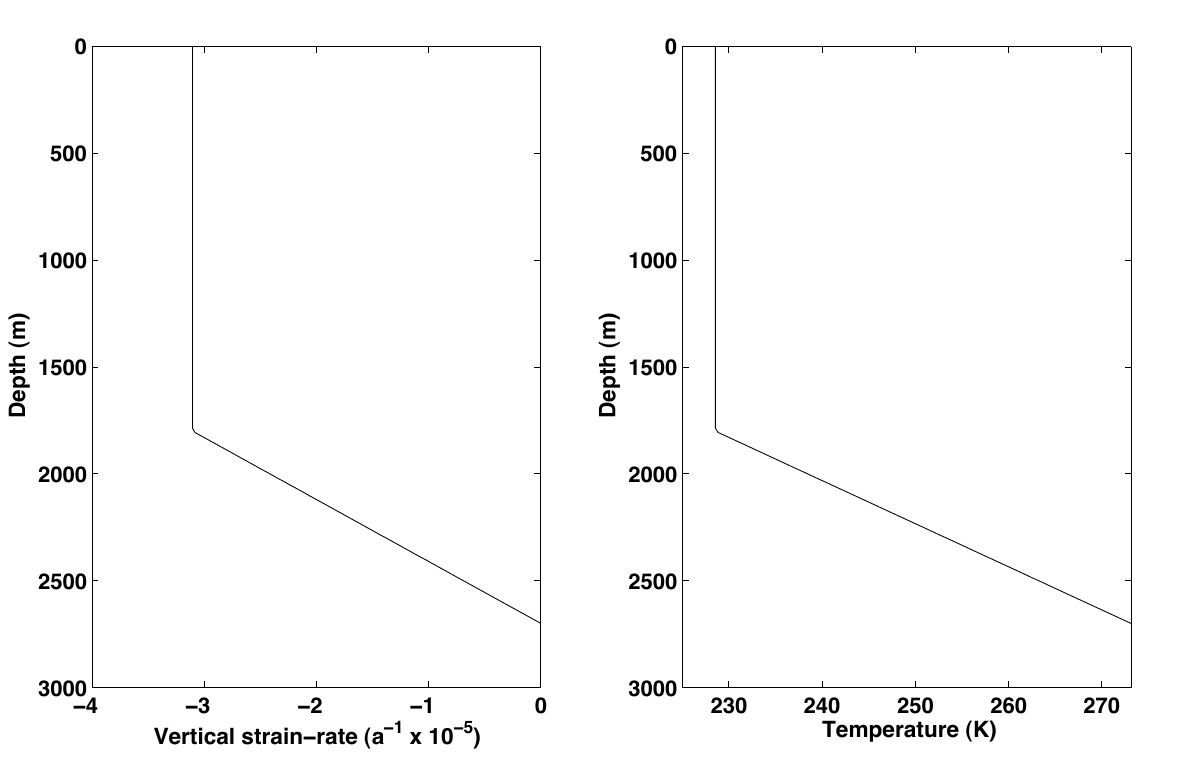}
  \caption{\emph{Dansgaard-Johnsen type distributions of
  the vertical strain rate \emph{(left panel)} and the temperature
  at the EDML site \emph{(right panel)}.
  The depth of the kinks is at two-thirds of the local ice thickness.
  The strain rate at the surface has been chosen such that the
  downward vertical velocity equals the accumulation rate, and
  the surface and basal temperatures match the ice-core data.}}
  \label{fig_dansgaard_johnsen}
\end{figure*}

According to the topographical data by Wesche et~al.\
\cite{wesche_etal_07}, the $x$-axis is approximately aligned
with the downhill direction, and the gradient of the free
surface elevation, $h$, is
\beq
  \pabl{h}{x} = -9 \times 10^{-4} \pm 10\%,
  \quad
  \pabl{h}{y} = 0.
\eeq
Thus, in the shallow-ice approximation, the only non-vanishing
bed-parallel shear-stress component is $T_{xz}$
\linebreak
($=S_{xz}$), given by
\beq
  T_{xz} = S_{xz} = \rho g z \pabl{h}{x},
  \qquad
  \label{eq_txz}
\eeq
where $g$ is the acceleration due to gravity. Combination with
the $x$-$z$-component of the Glen's flow law
(\ref{eq_glens_flow_law}) yields the isotropic horizontal
velocity,
\beq
  v_x = -2\rho g \pabl{h}{x}
         \int\limits_{z}^{H}
         A(T')\,\sigma^{n-1}\,\bar{z}\,\D\bar{z}
  \label{eq_vxy_iso}
\eeq
(e.g., Greve \cite{greve_97a}, Greve and Blatter
\cite{greve_blatter_09}), where $H$ is the ice thickness, the
rate factor $A(T')$ and stress exponent $n$ are chosen as
listed in Sect.~\ref{sect_glen}, and the enhancement factor $E$
has been set to unity. Similarly, for anisotropic conditions
and the corresponding flow law (\ref{eq_glens_flow_law_aniso}),
the horizontal velocity is
\beq
  v_x = -2\rho g \pabl{h}{x}
         \int\limits_{z}^{H}
         \hat{E}(\mathcal{A})\,A(T')
         \,\sigma^{n-1}\,\bar{z}\,\D\bar{z},
  \label{eq_vxy_aniso}
 \eeq
with the enhancement factor function $\hat{E}(\mathcal{A})$ of
Eq.~(\ref{eq_aniso_enh_factor}). Note that no-slip conditions
have been assumed at the ice base, that is, $v_x(z\!=\!H)=0$.

The unknowns in Eq.~(\ref{eq_vxy_aniso}) are the normal
deviatoric stresses ($S_{xx}$, $S_{yy}$, $S_{zz}$) which are
required together with the shallow-ice shear stress
(\ref{eq_txz}) for computing the deformability $\mathcal{A}$ by
Eq.~(\ref{eq_deformability}), and then the enhancement factor
$\hat{E}(\mathcal{A})$ by Eq.~(\ref{eq_aniso_enh_factor}). The
normal deviatoric stresses are computed by application of the
inverse anisotropic flow law
(\ref{eq_glens_flow_law_aniso_inv}) with the deformability in
the form (\ref{eq_deformability_inv}). The latter is evaluated
with the calculated shallow-ice deformations and an assumed
vertical strain rate $D_{zz}$ in the form of a
Dansgaard-Johnsen type distribution
\cite{dansgaard_johnsen_69}, which consists of a constant value
of $D_{zz}$ from the free surface down to two thirds of the ice
thickness, and a linearly decreasing value of $D_{zz}$ below. A
similar distribution is employed for the temperature profile
(see Fig.~\ref{fig_dansgaard_johnsen}). We also assume
extension in the $x$-direction only, so that the only non-zero
horizontal strain rate entering the evaluation of
Eq.~(\ref{eq_deformability_inv}) is $D_{xx}=-D_{zz}$. The
vertical velocity $v_z$ results from integrating the prescribed
vertical strain rate $D_{zz}$, which gives a linear/quadratic
profile (e.g., Greve et~al.\ \cite{greve_etal_02}).

For the ODF, we use the preliminary data of the EDML fabric
described above. However, since during the drilling process the
orientation of the core is not fixed, the horizontal
orientation of the non-circularly symmetric girdle fabric
(between approximately 600 and 2040~m depth) relative to our
coordinate system, i.e., the direction of ice flow, is unknown.
For this reason, we need to assign an orientation for the
fabric when computing the enhancement factor. We consider two
limiting cases by rotating the initial data such that the
girdle fabric at all depths is aligned with the $x$-axis (case
``R13'') and with the $y$-axis (case ``R23''), respectively.
This is illustrated in Fig.~\ref{fig_fabric_rotate}.

\begin{figurehere}
  \medskip
  \centering
  \noindent\includegraphics[width=52mm]{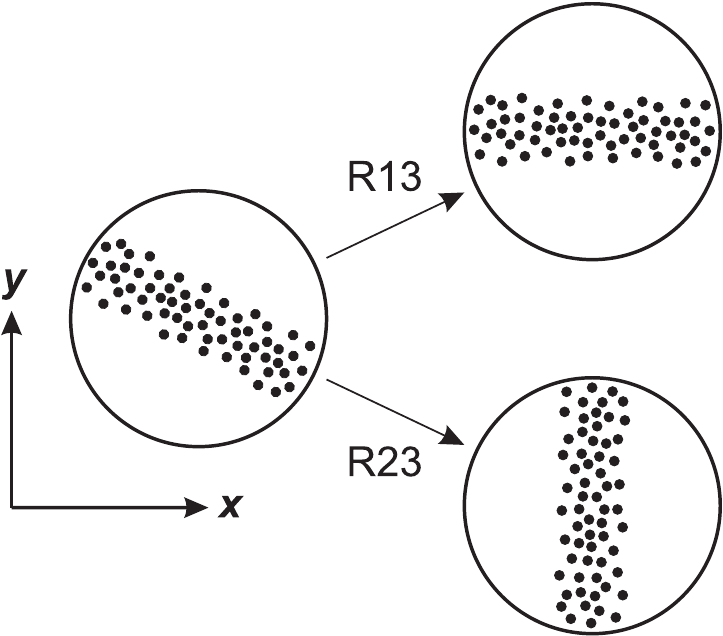}
  \caption{\emph{Sketch of the rotation of the girdle fabrics in
  order to align with the $x$-axis (case ``R13'') and with the
  $y$-axis (case ``R23'') in the Schmidt projection.}}
  \label{fig_fabric_rotate}
  \medskip
\end{figurehere}

At the surface, we assume isotropic conditions, so that
$\mathcal{A}_\mathrm{s}=1$, and for the maximum softening and
hardening parameters, we use the values $E_\mathrm{max}=10$ and
$E_\mathrm{min}=0$, respectively.

In a second step, we attempt at solving the fabric evolution
equation (\ref{eq_omb}). For the lack of better knowledge, we
neglect recrystallization, that is, we set $\lambda=0$ and
$\Gamma=0$ in the constitutive relations (\ref{eq_ofl}) and
(\ref{eq_opr}), respectively. By allowing only a dependency of
the orientation mass density $\rho^{\star}$ on the vertical
coordinate $z$ (one-dimensional steady-state problem) and on
the orientation $\vecbu{n}$, the orientation mass balance
(\ref{eq_omb}) yields an equation which governs the fabric
evolution along the EDML ice core,
\beq
  \pabl{\rho^{\star}}{z} v_{z} + \pa_{i}(\rho^{\star}u_{i}^{\star}) = 0,
  \label{eq_omb_1d_1}
\eeq
where the orientational gradient operator $\pa_{i}$ and the
orientation transition rate $u_{i}^{\star}$, respectively, read
in index notation as
\beq
  \pa_{i} = \pabl{}{n_{i}} - n_{i}n_{j} \pabl{}{n_{j}}
  \label{div_op_index}
\eeq
and, due to Eq.~(\ref{eq_otr}),
\beq
  u^{\star}_{i} = \iota D_{hk}n_{h}n_{k}n_{i}-\iota
  D_{ij}n_{j}+W_{ij}n_{j}.
  \label{eq_otr_index}
\eeq
With Eq.~(\ref{div_op_index}), and by inserting the
constitutive relation Eq.~(\ref{eq_otr_index}) in
Eq.~(\ref{eq_omb_1d_1}), it follows
\beq
  \begin{array}[b]{l}
    \bigpabl{\rho^{\star}}{z} v_{z}
    + u_{i}^{\star}\pa_{i}\rho^{\star}
    + \rho^{\star}\pa_{i}u_{i}^{\star}
    \\[2.5ex]
    = \bigpabl{\rho^{\star}}{z} v_{z}
      + (W_{ij}-\iota D_{ij})n_{j}\pa_{i}\rho^{\star}
    \\[2ex]
    \hspace*{14mm}
      +\;3 \iota \rho^{\star}D_{hk}n_{k}n_{h}
    = 0.
  \end{array}
  \label{eq_omb_1d_2}
\eeq
We assume that the local flow field consists of vertical
compression with the compression rate (negative vertical strain
rate) $\varepsilon=-\pa{}v_{z}/\pa{}z$ according to
Fig.~\ref{fig_dansgaard_johnsen}, horizontal extension in
$x$-direction, and the horizontal, bed-parallel shear rate
\beq
  \gamma = \pabl{v_x}{z}
         = 2 \rho g \pabl{h}{x}\,
             \hat{E}(\mathcal{A})\,A(T')\,\sigma^{n-1}\,z
  \label{eq_shear_rate_aniso}
\eeq
that results from Eq.~(\ref{eq_vxy_aniso}). The velocity
gradient $\tenssu{L}=\grad\vecbu{v}$ then reads
\beq
  \tenssu{L} = \matdd{\varepsilon}{0}{\gamma}
                     {0}{0}{0}
                     {0}{0}{-\varepsilon}.
\eeq
Consequently, we obtain for the strain-rate tensor $\tenssu{D}$
and the spin tensor $\tenssu{W}$
\beq
  \tenssu{D} = \matdd{\varepsilon}{0}{\hf{}\gamma}
                     {0}{0}{0}
                     {\hf{}\gamma}{0}{-\varepsilon}
\eeq
and
\beq
  \tenssu{W} = \matdd{0}{0}{\hf{}\gamma}
                     {0}{0}{0}
                     {-\hf{}\gamma}{0}{0}.
\eeq
With these expressions and the introduction of spherical
coordinates, Eq.~(\ref{eq_omb_1d_2}) reduces to
\beq
  \begin{array}[b]{l}
    4 \bigpabl{\rho^{\star}}{z} v_{z}
    \\[2.5ex]
     +\;3 \iota \rho^{\star}
        \big[ \varepsilon
          (2\sin^{2}{\theta}\cos{2\varphi}-1-3\cos{2\theta})
    \\[2.5ex]
    \hspace*{11mm}
     +\;2 \gamma \sin{2 \theta} \cos{\varphi} \big]
    \\[2ex]
     +\;2 \bigpabl{\rho^{\star}}{\varphi}
      \Big[\varepsilon \iota \sin{2\varphi}
     +\;\gamma(-1+\iota)\,\bigfrac{\sin{\varphi}}{\tan{\theta}}
      \Big]
    \\[2.5ex]
     +\;2 \bigpabl{\rho^{\star}}{\theta}
        \Big[ -\bigfrac{1}{2}\iota \varepsilon
        \sin{2\theta}\,(\cos{2\varphi}+3)
    \\[2.5ex]
    \hspace*{14.5mm}
     +\;\gamma(1-\iota \cos{2\theta}) \cos{\varphi}
     \Big]
     = 0,
  \end{array}
  \label{eq_omb_1d_3}
\eeq
where $\theta$ and $\varphi$ are the polar angle (co-latitude)
and azimuth angle (longitude), respectively. Note that, due to
Eq.~(\ref{eq_shear_rate_aniso}), the shear rate $\gamma$
depends on the fabric via the deformability $\mathcal{A}$.

\begin{figure*}[htb]
  \centering
  \includegraphics[width=165mm]{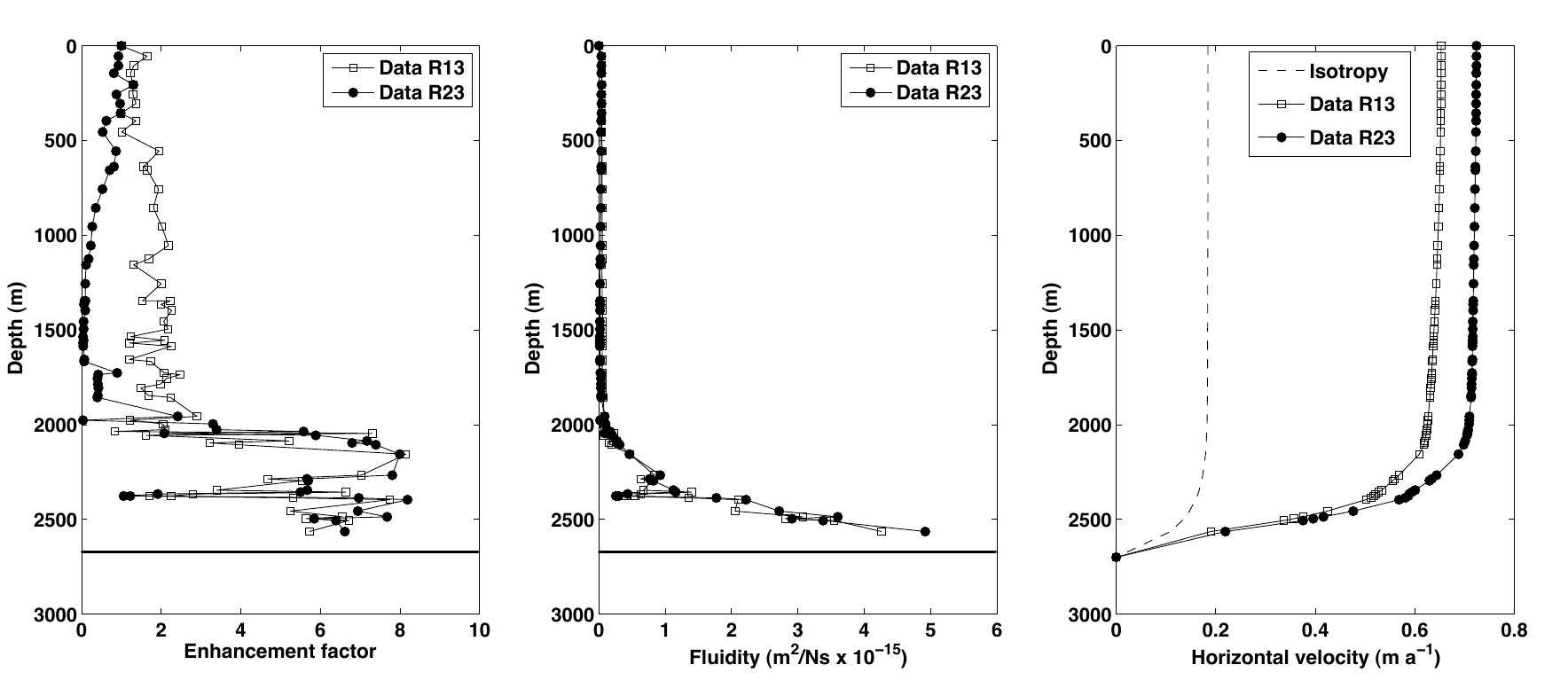}
  \caption{\emph{Variation of
  the enhancement factor \emph{(left panel)},
  the ice fluidity \emph{(middle panel)} and
  the horizontal velocity \emph{(right panel)}
  along the EDML ice core.
  ``Data R13'' and ``Data R23'' represent the solutions
  obtained with the measured girdle fabrics rotated to align
  with the $x$- and $y$-direction, respectively, and
  ``Isotropy'' represents isotropic conditions.}}
  \label{fig_enh_flui_vel}
\end{figure*}

The shear flow at the EDML station leads to the transport of
ice particles over significant horizontal distances. Huybrechts
et~al.\ \cite{huybrechts_etal_07} estimate, based on
three-dimensional flow modelling, that particles at 89\% depth
of the core originate from $\approx{}184\,\mathrm{km}$
upstream. This is not taken into account in our spatially
one-dimensional model. However, the variation of the shear
upstream of the drill site is likely small due to the small
variation of the surface gradient
(Fig.~\ref{fig_coordinate_system}), so that the error resulting
from the neglected horizontal inhomogeneity should be limited.

In this study, we restrict the solution of
Eq.~(\ref{eq_omb_1d_3}) to the simplified case of a
transversely isotropic (circularly symmetric) fabric, so that
the OMD $\rho^{\star}$ is only a function of the depth $z$ and
the polar angle $\theta$. Then Eq.~(\ref{eq_omb_1d_3}) becomes,
after integration over the azimuth angle $\varphi$,
\beq
  \begin{array}[b]{l}
    4 \bigpabl{\rho^{\star}}{z} v_{z}
    - \bigpabl{\rho^{\star}}{\theta} 3 \iota \varepsilon \sin{2\theta}
    \\[2.5ex]
    \hspace*{12mm}
    -\;3 \iota \rho^{\star}\varepsilon (1+3\cos{2\theta}) = 0.
  \end{array}
  \label{eq_tranverse}
\eeq
Equation (\ref{eq_tranverse}) is solved by using a
finite-difference discretization with the parameter
$\iota=0.6$.

\subsection{Results}

Figure~\ref{fig_enh_flui_vel} shows the variation of the
enhancement factor, the ice fluidity and the horizontal
velocity along the ice core, computed with the ODF based on the
fabric data described above. For both limiting cases R13 and
R23, the enhancement factor is close to unity in the upper
600~m, which reflects the nearly isotropic fabrics in that part
of the EDML core. Further down, in the girdle fabric regime,
the case R13 is characterized by a moderate increase of the
enhancement factor to an average value of about two, whereas
the case R23 exhibits a strong decrease of the enhancement
factor to values close to zero. This demonstrates clearly that
the girdle fabrics produce a significantly different mechanical
response depending on the orientation relative to the ice flow.
Case R23 is probably closer to reality, because in the girdle
fabric regime above 2000~m depth the deformation is essentially
pure shear (vertical compression, horizontal extension in
$x$-direction only). For this situation, a simple
``deck-of-cards'' model illustrates that the $c$-axes turn away
from the $x$-axis and towards the $z$-axis, whereas nothing
happens in $y$-direction, so that in the Schmidt projection a
concentration perpendicular to the $x$-axis (flow direction)
results.

\begin{figure*}[htb]
  \centering
  \includegraphics[width=110mm]{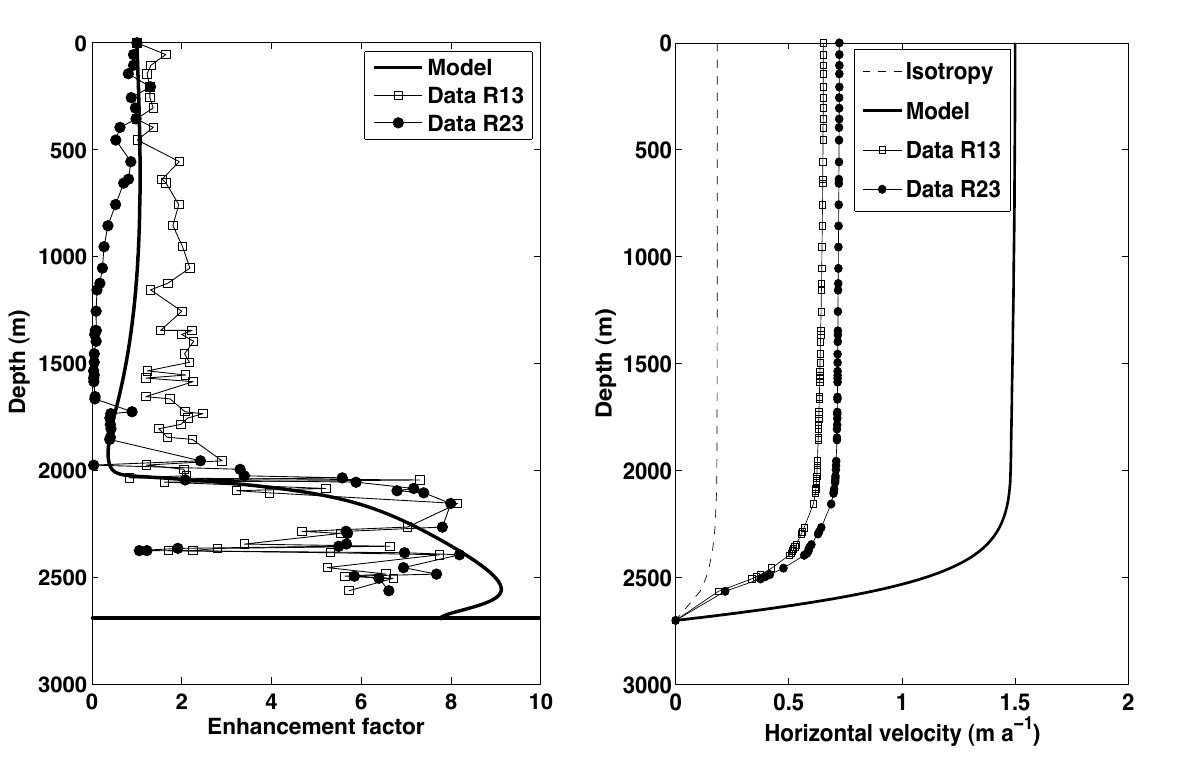}
  \caption{\emph{Variation of
  the enhancement factor \emph{(left panel)} and
  the horizontal velocity \emph{(right panel)}
  along the EDML ice core.
  ``Model'' represents the solutions based on the fabric
  evolution equation (\ref{eq_tranverse}) for transverse
  isotropy. For ``Data R13'', ``Data R23'' and ``Isotropy''
  see the caption of Fig.~\ref{fig_enh_flui_vel}.}}
  \label{fig_transverse}
\end{figure*}

Below 2000~m depth, where the fabric switches to a single
maximum, the difference between the cases R13 and R23
essentially vanishes. The crystallite basal planes are
favourably oriented for the now prevailing simple-shear
deformation, which leads to large deformabilities.
Consequently, the enhancement factor shows a sharp increase to
a maximum value of about eight, which is close to the
theoretical maximum of $E_\mathrm{max}=10$.

The variabilities of the enhancement factor and the effective
stress, as well as the increase of the temperature with depth,
contribute to the fluidity profiles shown in
Fig.~\ref{fig_enh_flui_vel}b. Since the fluidity is very small
above 2000~m depth and increases only further down, the
difference between the cases R13 and R23 in absolute values is
surprisingly small. At 2563~m depth, the fluidity is about 200
times higher than the fluidity at 1000~m depth for the case R23
due to the counteracting contributions from the favourably
oriented $c$-axes, the higher temperature and the smaller
effective stress. The latter is somewhat surprising; it is
caused by the normal deviatoric stresses $S_{xx}$ and $S_{zz}$,
which decrease strongly below 2000~m depth and outweigh the
influence of the increasing shear stress $S_{xz}$ in the
effective stress.

Owing to the large enhancement factors close to the bottom, the
anisotropic flow law predicts significantly larger horizontal
velocities compared to the isotropic flow law for the entire
depth of the ice core (Fig.~\ref{fig_enh_flui_vel}c). At the
surface, the anisotropic horizontal velocities are by
approximately a factor 3.5 larger than their isotropic
counterparts, and the absolute value of
$\approx{}0.7\,\mathrm{m\,a^{-1}}$ agrees very well with
measurements (H.~Oerter, pers.\ comm.\ 2005; Wesche et~al.\
\cite{wesche_etal_07}). The difference between the cases R13
and R23 amounts to $\approx{}10\%$, the larger values
being obtained for the case R23 owing to the slightly larger
enhancement factors below 2000~m depth. Interestingly, these
differences show that the fabrics are not perfectly
transversely isotropic below 2000~m depth, even though they are
very close to the single-maximum type.

Let us now turn to the simulation in which the fabric evolution
is computed by solving Eq.~(\ref{eq_tranverse}) for a
transversely isotropic fabric. Although this assumption is not
consistent with the observed girdle fabric between
approximately 600 and 2000~m depth and is therefore a gross
simplification, it is interesting to study the mechanical
response of such a simplified system and the differences to the
ice flow resulting from applying the measured fabrics.

Figure~\ref{fig_transverse}a shows the comparison between the
enhancement factors resulting from the computed, transversely
isotropic fabric (which will be referred to as ``modelled
enhancement factor'' in the following) and from the fabric
data. Evidently, the agreement is good despite the assumption
of transverse isotropy. Down to 1800~m depth, the modelled
enhancement factor lies in between the cases R13 and R23, which
are the limiting cases for the orientation of the measured
girdle fabric with respect to the ice-flow direction. Between
1800 and 1900~m depth, the modelled enhancement factor is very
close to the low values of the case R23, for which the girdle
fabric is aligned perpendicular to the flow direction. Below
2000~m depth, the sharp increase is also well reproduced;
however, the maximum of the modelled enhancement factor is more
pronounced and lies closer to the bottom than for the cases R13
and R23.

For that reason, the modelled enhancement factor leads to
larger near-basal shear rates than the enhancement factor based
on the cases R13 and R23. Consequently, the horizontal velocity
resulting from the modelled enhancement factor is larger by
about a factor two than the velocities for the cases R13 and
R23 (Fig.~\ref{fig_transverse}b). At the surface, a value of
$\approx{}1.5\,\mathrm{m\,a^{-1}}$ is reached, which is twice
the measured surface velocity. This highlights the great
sensitivity of the ice dynamics to the processes near the
bottom, which are most difficult to model precisely. Beside the
assumption of transverse isotropy, a weak point in that context
is the neglection of recrystallization processes, which are
expected to become important for the fabric evolution in the
lower part of the ice core. This point requires further
attention.

\section{Conclusions}

The newly developed CAFFE model (Continuum-mechanical,
Anisotropic Flow model, based on an anisotropic Flow
Enhancement factor), which comprises an anisotropic flow law as
well as a fabric evolution equation, was presented in this
study. It is a good compromise between physical adequateness
and simplicity, and is therefore well suited for being used in
flow models of ice sheets and glaciers.

The CAFFE model was successfully applied to the site of the
EDML ice core in East Antarctica. Two different methods were
employed, (i) computing the anisotropic enhancement factor and
the horizontal flow based on fabrics data, and (ii) solving the
fabric evolution equation under the simplifying assumption of
transverse isotropy. Method (i) demonstrated clearly the
importance of the anisotropic fabric in the ice column for the
flow velocity, and better agreement with the measured surface
velocity was achieved compared to an isotropic computation. The
anisotropic enhancement factor produced with method (ii) agreed
reasonably well with that of method (i), despite the fact that
the measured fabric is not transversely isotropic in large
parts of the ice core.

A solution of the fabric evolution equation (\ref{eq_omb_1d_3})
for the EDML ice core without the assumption of transverse
isotropy has been presented elsewhere (Seddik et~al.\
\cite{seddik_etal_08}). Further, the CAFFE model has already
been implemented in the three-dimensional, full-Stokes ice-flow
model Elmer/Ice (Seddik \cite{seddik_08}, Seddik et~al.\
\cite{seddik_etal_09a}) in order to simulate the ice flow in
the vicinity within $100\,\mathrm{km}$ around the Dome Fuji
drill site (Motoyama \cite{motoyama_07}) in central East
Antarctica.

\vspace*{1ex}

\begin{center}
\includegraphics[scale=0.33]{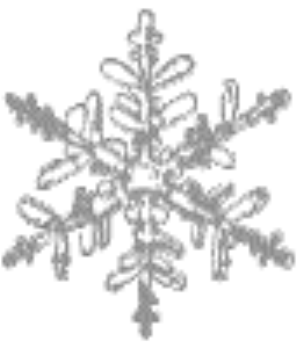}
\hspace*{2em}
\includegraphics[scale=0.33]{Snowflake.pdf}
\hspace*{2em}
\includegraphics[scale=0.33]{Snowflake.pdf}
\end{center}

\section*{Acknowledgements}

The authors wish to thank Dr.\ S\'ergio H.\ Faria (University
of G\"ottingen, Germany), Dr.\ Olivier Gagliardini (Laboratory
of Glaciology and Environmental Geophysics, Grenoble, France)
and Professor Kolumban Hutter (Swiss Federal Institute of
Technology, Zurich) for their collaboration in developing the
new model for anisotropic polar ice. Thanks are further due to
Ms.\ Ilka Hamann and  Dr.\ Sepp Kipfstuhl for kindly providing
the preliminary fabric data of the EDML ice core, to Dr.\ Hans
Oerter for communicating the measured surface velocity, to Ms.\
Christine Wesche (all at Alfred Wegener Institute for Polar and
Marine Research, Bremerhaven, Germany) for allowing us to use
the topographic map of the vicinity of the drill site, and to
an anonymous reviewer whose comments helped improving the
clarity of the paper.

This work was supported by a Grant-in-Aid for Creative
Scientific Research (No.\ 14GS0202) from the Japanese Ministry
of Education, Culture, Sports, Science and Technology, and by a
Grant-in-Aid for Scientific Research (No.\ 18340135) from the
Japan Society for the Promotion of Science. We would like to
express our gratitude for the efficient management of the
Creative Research project by the leader, Professor Takeo
Hondoh, and the project assistant, Ms.\ Kaori Kidahashi.

\clearpage



\end{multicols}
\end{document}